Commentary

# AI incidents and 'networked trouble': The case for a research agenda

Tommy Shaffer Shane

## Abstract
Against a backdrop of widespread interest in how publics can participate in the design of AI, I argue for a research agenda focused on AI incidents – examples of AI going wrong and sparking controversy – and how they are constructed in online environments. I take up the example of an AI incident from September 2020, when a Twitter user created a 'horrible experiment' to demonstrate the racist bias of Twitter's algorithm for cropping images. This resulted in Twitter not only abandoning its use of that algorithm, but also disavowing its decision to use *any* algorithm for the task. I argue that AI incidents like this are a significant means for participating in AI systems that require further research. That research agenda, I argue, should focus on how incidents are constructed through networked online behaviours that I refer to as 'networked trouble', where formats for participation enable individuals and algorithms to interact in ways that others – including technology companies – come to know and come to care about. At stake, I argue, is an important mechanism for participating in the design and deployment of AI.

This article is a part of special theme on Analysing Artificial Intelligence Controversies. To see a full list of all articles in this special theme, please click here: https://journals.sagepub.com/page/bds/collections/analysingartificialintelligencecontroversies

In September 2020, a Twitter user posted a tweet that he described as 'a horrible experiment'. He tweeted two long images, with former U.S. president Barack Obama at one end, and Republican senator Mitch McConnell at the other. The tweet (shown in Figure 1) presented an implied test of Twitter's algorithm for cropping images, as it would need to 'pick' which face to centre and which to crop out to make the images fit within the tweet. In both images, the algorithm cropped out the face of Obama and centred the face of McConnell. The tweet went viral, amplifying accusations of racist bias around the world, and sparking a meme into being through which many other people sought to 'test' the algorithm. Soon after, Twitter announced it would abandon the cropping algorithm, claiming not just that the algorithm was harmful, but that they were wrong to use any algorithm for deciding whose faces are visible on their platform (Chowdhury, 2021). Against a backdrop of widespread interest in how publics can participate in and transform AI systems (Birhane et al., 2022; Bunz, 2022), this 'horrible experiment' suggests that interacting with algorithms – for example, testing them – may have the capacity to alter the design of algorithms and their influence on society.

In this short commentary, I argue for a research agenda focused on AI incidents like this one – defined as examples of AI going wrong in deployment and acquiring political capacities to alter the way that AI system works – and how they enable participation in AI. I take up the controversy over Twitter's cropping algorithm to argue that such a research agenda must pay attention to how individuals and algorithms interact, how those interactions are circulated and framed as incidents in online environments,

Digital Humanities Department, King's College London, London, UK

**Corresponding author:**
Tommy Shaffer Shane, Digital Humanities Department, King's College London, London WC2R 2LS, UK.
Email: tommy.shane@kcl.ac.uk

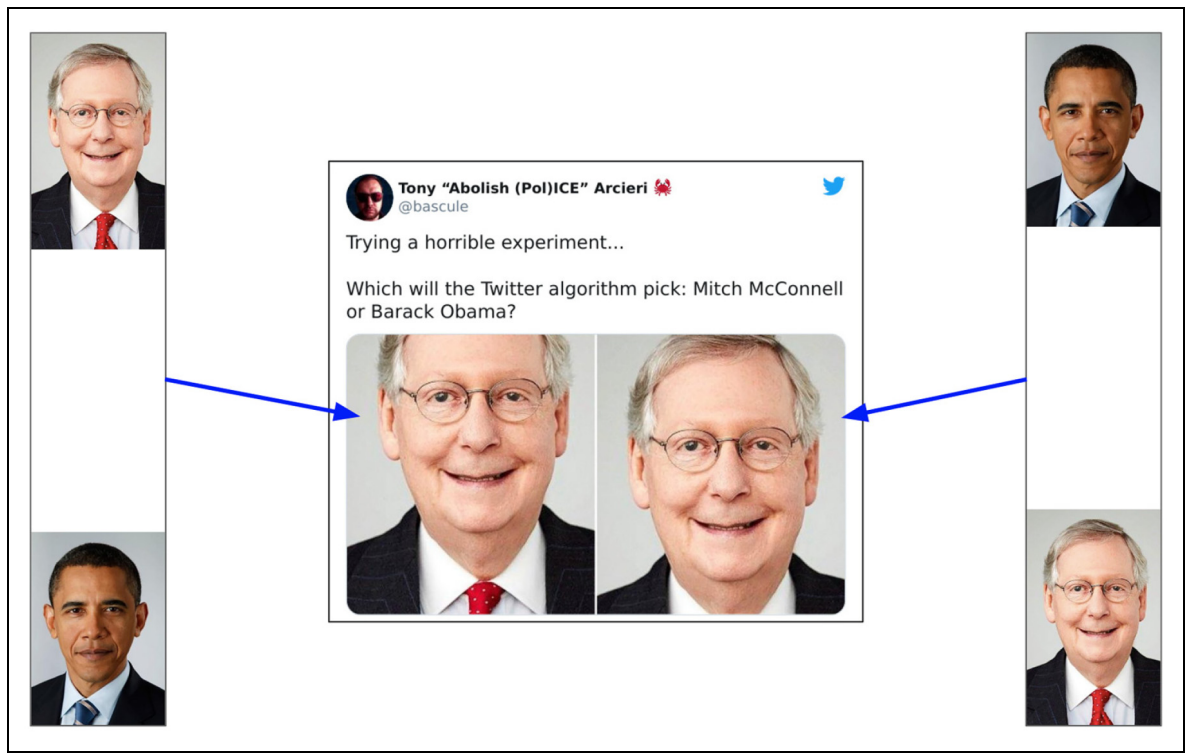

**Figure 1.** Visual explainer of how the Twitter user created a format to test the cropping algorithm.

and how others (e.g., online publics and technology companies) come to know and come to care, so that changes are made to the AI system.

To develop this argument, I first propose we understand the 'tests' of Twitter's cropping algorithm as an example of 'algorithm trouble': 'everyday encounters with artificial intelligence [that] manifest, at interfaces with users, as unexpected, failing, or wrong event[s]' (Meunier et al., 2021). However, I further develop the notion of 'trouble' by asking how trouble is *made*; how deliberate interactions with algorithms in deployment cause trouble for technology companies. To do so, I use Ahmed's (2017) theorisation of troublemaking, which provides an account of how actors can *make trouble* within social interactions to resist harmful social structures by deliberately causing breakdown. Second, I apply Ahmed's theorisation of troublemaking to a close reading of the Twitter cropping AI incident, arguing that it is better understood as 'networked trouble' due to the role of networked environments (such as social media) in making trouble. Of particular importance for 'networked trouble', I argue, is the discovery of a *format*, an adaptable method of interacting with an algorithm that can spread across networked environments and demonstrate a harm to which technology companies feel obliged to respond. Finally, I close by suggesting that attending to 'networked trouble' is necessary for analyses of how AI incidents form in today's online environments, an increasingly prominent method of participating in AI systems.

## Trouble and troublemaking

### Trouble

In recent years, examples of AI going wrong and sparking controversy – what I am referring to as AI incidents – have become both widespread and influential. AI incidents have included an AI model offensively labelling a photo of an African-American couple as 'gorillas', and Google translating the genders of job roles (e.g., doctor and nurse) in terms of sexist stereotypes (Crawford, 2017). Thousands of incidents have been documented through initiatives such as the AI Incident Database (2023) and have been used to critique algorithms' design and underlying logic (e.g., Benjamin, 2019), often resulting in successful pressure on technology companies to alter their AI systems.

Meunier et al. (2021) propose we think of AI incidents as 'algorithm trouble': 'unsettling events that may elicit, or even provoke, other perspectives on what it means to live with algorithms'. Their framing suggests that AI incidents are not simply a question of AI going wrong but also individuals and institutions coming to be *troubled* by AI going wrong. It is the troubling nature of AI incidents – not just the fact of their technical malfunction – that coalesces a public with the capacity to force technology companies into action. AI incidents therefore involve an articulation of the relevance of issues (e.g., racial bias) and are thus relevant to an understanding of the formation of publics (Marres, 2012). As such, precisely *how* algorithms come to trouble actors is a salient feature of an analysis of AI incidents. AI incidents, and how they trouble publics and technology companies, therefore offer important empirical occasions for understanding how AI systems can be transformed.

In this commentary, I argue that trouble can be further elaborated as *troublemaking*, on the basis that trouble is not found, but *made*. There is a process by which publics transform 'seemingly simple quirks and individually felt glitches into shared social consequences with the power to shape social life' (Ananny, 2022, 5). AI incidents, I argue, are constructed by networked behaviours, and therefore provide us with valuable case studies for understanding AI's processes of issue articulation. It is through a detailed analysis of these case studies – how AI going wrong acquires political capacities through networked behaviours – that AI incidents' capacity for changing AI systems can be understood. To do this, I propose we draw on Sara Ahmed's theory of how social interactions can be disrupted to *make* trouble.

## Troublemaking

Ahmed takes up troublemaking as a central concept in her feminist theory (Ahmed, 2017) to describe how interactions can be deliberately disrupted to critique and challenge harmful systems. Her account is helpful for an understanding of AI incidents because it offers a theoretical framework for how individuals can deliberately controversialise elements of social life such that they are experienced as troubling by others and in need of remedy. As a result, this theoretical account can provide questions to ask as part of an analysis of how AI incidents are *made*, which I will elaborate on in this section.

Ahmed presents three techniques for troublemaking: pointing out invisible structures, orientating actors around them, and problematising the harm they cause. The first technique – pointing – involves bringing features of a social environment into view, specifically those that are usually invisible, in order that they can be challenged. In Ahmed's (2017, 255) words, some people are 'bruised by structures that are not even revealed to others'. Troublemaking involves pointing out those ignored instances of harm in social settings to direct attention to them (Ahmed, 2017). This could include pointing out that a joke is sexist, or that someone is being discriminatorily excluded. 'Making feminist points, antiracist points, sore points,' Ahmed writes, 'is about pointing out structures that many are invested in not recognizing' (Ahmed, 2017, 158). It is by pointing out those invisible structures that others can come to be troubled cby them. From this insight, we can take a question for AI incidents: *How do actors point out invisible features of algorithms?*

The second technique involves achieving a shared orientation. For Ahmed, trouble is experienced when actors become collectively oriented around a harmful element of a social environment (such as a sexist joke) and achieve a shared experience. Often this relies on emotive provocation. By focusing attention on harmful features of social life, a troublemaker can disturb or possibly outrage those present to acquire their collective attention. This can lead to interactions breaking down: previously people were laughing at the joke's punchline; now they're feeling uncomfortable about its sexism. Provoking breakdown offers a technique for bringing actors into a shared experience in which issues can be articulated and debated. This is not to say that those present agree with each other, but that *they agree on the points on which they disagree*. This technique offers a second question for AI incidents: *How do actors achieve a shared orientation towards an algorithm and its features?*

The third technique is problematising harm. Ahmed's account of troublemaking is specifically interested in exposing harms, with the act of making trouble being designed to contest discriminatory social structures. In this respect, Ahmed translates Garinkel's ethnomethodology into a political tactic. Garfinkel famously taught his students to '[s]tart with familiar scenes and ask what can be done to *make trouble*' (1967, 37; italics mine). For Garfinkel, troublemaking was a way of forcing social interactions into breakdown in order to make visible the assumptions and agreements that underpin sociality, and therefore as a technique of social inquiry. For Ahmed, troublemaking forces social interaction into breakdown in order to frame it as harmful and in need of remedy, and therefore as a technique of political action. From this, we can draw a third question for AI incidents: *How do actors frame features of algorithms as harmful?*

To sum up, troublemaking is a technique for pointing out and orienting actors around harmful structures by deliberately disrupting interactions. And it prompts us to ask of AI incidents: How did actors point out something about an algorithm that was previously invisible? How did actors use breakdown to achieve a shared orientation towards it? And how did they frame it as harmful? In the following section, I show how these questions can be useful for analysing AI incidents that unfold in networked environments such as social media, by applying them to Twitter's image cropping AI incident.

## Networked trouble

In this section, I suggest that the 'horrible experiment' (Figure 1) can be understood as a form of troublemaking. Drawing on Ahmed, I ask: how did actors point, mutually orientate, and problematise harm? I argue that because these interactions unfolded in networked environments, meaning actors are connected across disparate spaces and times rather than in the same room (Cetina, 2009), trouble unfolded in particular ways. It relied on networked media, a format for participation, and the production of inscriptions. These features, which I elaborate in this section, point towards characteristics of troublemaking that unfolds in networked environments such as social media, or what I am calling 'networked trouble'.

### How did actors point?

The first observation we can make about the cropping AI incident is that it required the networking of media. The Twitter user networked his tweet with his followers; his followers networked it with their followers through likes and retweets and memetic remixes; and journalists embedded screenshots of them in their news stories. The troublemaking was networked, then, in the sense that it depended on networking media (photos, tweets, screenshots and news articles) with audiences to bring the previously invisible cropping algorithm into public awareness. This 'networked pointing', where people pointed at the algorithm's unrecognised biases through networked media, was necessary for coordinating the attention of actors that were fragmented across the spaces and times of digital environments. This in turn set the stage for disagreement and intervention by leveraging the affordances of networked environments to bring the previously invisible algorithm and its stakes into view.

### How did actors achieve a shared orientation?

Shared orientation was made possible by the Twitter user's discovery of a *format* for troublemaking, whereby Twitter users and the algorithm could interact to make trouble. Formats enable actors to participate in an activity (Kelty, 2019), and this is what the Twitter user discovered with his 'horrible experiment': a format – two long images with faces readable as different races on either end – in which he and the algorithm could coordinate to demonstrate the racist bias. The format enabled him and the algorithm to both be orientated by the same elements of the images (the faces) and interact to produce a reproducible effect. In this respect, the *algorithm* was invited to participate in the 'horrible experiment'. In addition, other users were also invited to participate, due to the establishment of a meme format. Many remixed the format, such as using Simpsons characters (shown in Figure 2), and combined tonal registers (serious moral claims and playful jokes) to articulate the relevance of the issue to a larger and more diverse public. These remixes enabled an online public to emerge across a fragmented networked environment, all oriented around the algorithm. What's more, the framing of the format as a 'test' used the technology industry's own vernacular as a (successful) claim for its attention. The consequences of the format's ease of use and adaptability indicate that networked trouble may require *a format* for participation. One further implication is that harmful algorithms, given the right format, may themselves be participants in contesting their own design and deployment.

### How did actors problematise harm?

Central to the "horrible experiment" was the production of an *inscription* (Latour, 1987): the tweets that presented the cropping algorithm's harmful behaviour. These inscriptions possessed certain qualities. The first is that the inscriptions were regarded as providing *evidence* of the algorithm's bias because they were a record or imprint of its behaviour. Second, the meaning of the inscriptions (i.e., the tweets) was durable in multiple online contexts: they could be screenshotted in news articles and retain their meaning even when out of context. Third, the inscriptions framed the algorithm as racist, providing a severe claim of harm that attributed the racism to the algorithm (rather than a more complex sociotechnical assemblage), a technique of attribution that helped make the case for algorithm's redesign or removal. This type of inscription – evidence of the algorithm's behaviour that could be durably decontextualised and straightforwardly read as harmful – was important to the viral spread of the 'horrible experiment' and its political capacities. These qualities suggest that networked trouble problematises harm through particular kinds of *inscriptions* produced by interactions with algorithms.

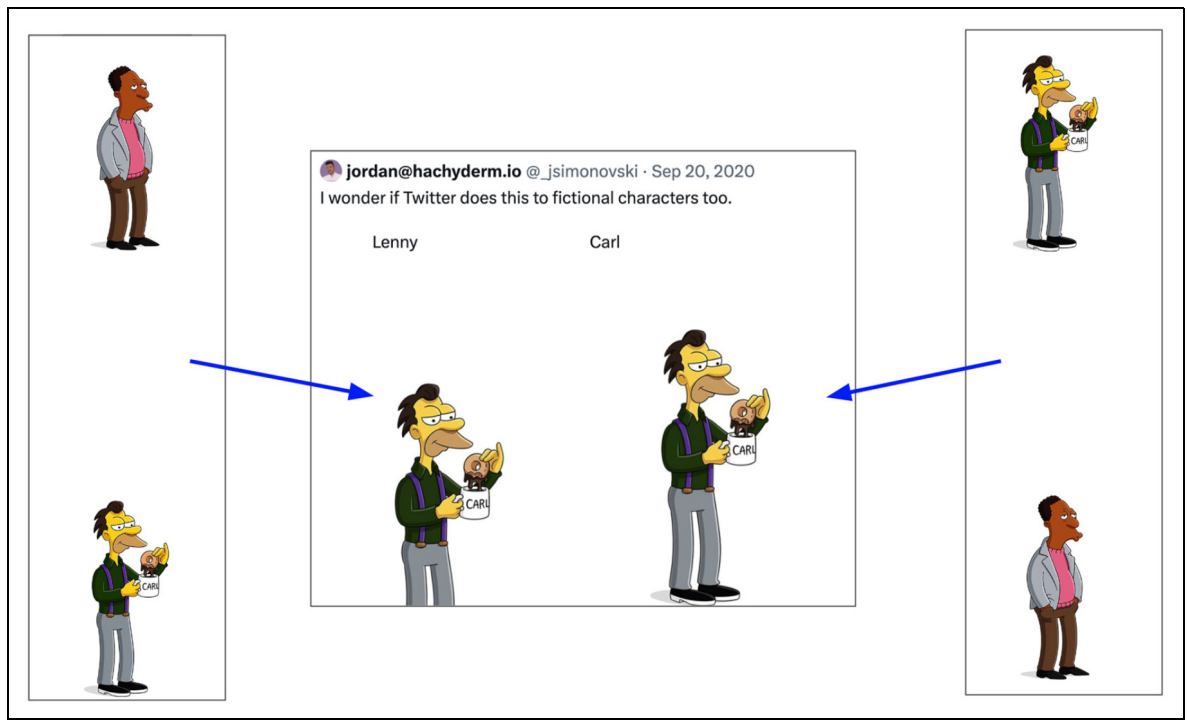

**Figure 2.** A visual explainer of the format for trouble applied to two characters from the cartoon TV series The Simpsons.

In sum, I have presented three hypotheses for the study of 'networked trouble' in AI incidents: that it relies on i) the networking of media and audiences, ii) a *format* in which diverse actors, including the algorithm itself, can participate, and iii) the production of inscriptions that can convincingly and durably present algorithmic behaviour as harmful. These hypotheses can aid, and are important for, an analysis of how AI incidents are constructed in networked environments.

## Networked trouble in AI incidents

To close, I argue that we require a research agenda to understand *how AI incidents are constructed through networked online behaviours* and suggest that understanding formats for trouble and how they materialise in online environments is a vital component of that agenda. To return to our case, at stake in the 'horrible experiment' was the discovery of a format for an interaction with the algorithm that could inscribe a harm to which a technology company felt compelled to respond. Understanding this field of concern, therefore, requires a research agenda that focuses on historical AI incidents and asks how formats materialise in networked environments such as social media and produce compelling inscriptions of harm. By focusing on these formats, empirical attention can turn from treating algorithmic 'bugs' as technical accidents that are simply found, to how we come to be troubled by AI going wrong, and therefore how publics form and acquire the capacity to alter algorithms' design and use. This can also allow researchers to interrogate the role of networked online environments (such as Twitter) in public participation in AI, and its limitations.

As a final reflection, I suggest it can be helpful to think of AI as a participant in its own transformation, because it plays an important role in generating inscriptions that demonstrate its potential harms. This has important consequences for institutional and activist efforts to enable participation in the development of AI. AI may be an ally in its own contestation or refusal, as actors are increasingly coordinating with harmful algorithms outside of institutionally managed exercises, using networked formats such as 'tests'. We might usefully consider AI as *a troublemaker* and pay more attention to the conditions that enable algorithms to make trouble for technology companies. And given that troublemaking with AI is arguably on the rise – one need just look at attempts to 'jailbreak' Chat-GPT (Xiang, 2023) to produce problematic outputs – so too, I suggest, is the urgency to stay with the networked trouble in AI incidents.

### Acknowledgements

I am very grateful to Professor Noortje Marres and Dr. Jonathan Gray for their helpful feedback on early drafts of this paper.

### Declaration of conflicting interests

The author declared no potential conflicts of interest with respect to the research, authorship, and/or publication of this article.

### Funding

The author disclosed receipt of the following financial support for the research, authorship, and/or publication of this article: This work was supported by the Arts and Humanities Research Council.

### ORCID iD

Tommy Shaffer Shane  https://orcid.org/0000-0001-9665-4284